\documentclass[letterpaper,10pt]{article}

\usepackage[utf8]{inputenc}
\usepackage{graphicx}
\usepackage{amsmath}
\usepackage{amssymb}
\usepackage{psfrag}

\newcommand{\dd}{\textrm{d}}

\newcommand{\hypergf}{{}_{2}F_{1}}

\title{New conditionally exactly solvable potentials of exponential type}

\author{A.\ L\'opez-Ortega \\
Departamento de F\'{\i}sica.\\ 
Escuela Superior de F\'{\i}sica y Matem\'aticas. \\
Instituto Polit\'ecnico Nacional. \\
Unidad Profesional Adolfo L\'opez Mateos. Edificio 9. \\
M\'exico, D.\ F., M\'exico. \\
C.\ P.\ 07738 \\
email: alopezo@ipn.mx }

\begin{document}

\maketitle

\begin{abstract}

Based on a method that produces the solutions to the Schr\"odinger equations of partner potentials, we give two conditionally exactly solvable partner potentials of exponential type defined on the half line. These potentials are multiplicative shape invariant and each of their linearly independent solution includes a sum of two hypergeometric functions. Furthermore we calculate the scattering amplitudes and study some of their properties.

KEYWORDS: Exactly solvable potentials; Shape invariance;  Hypergeometric function; Scattering amplitude.

PACS: 03.65.Ge, 03.65.Nk, 03.65.Ca, 02.90.+p

\end{abstract}

\section{Introduction}
\label{s: Introduction}

We know that in physics the exactly solvable problems are useful in the analysis of physical systems since they allow us to study in detail their properties or they are suitable approximations to more complex systems. In non relativistic quantum mechanics the potentials for which we can solve exactly the Schr\"odinger equation are used in the analysis of several phenomena. Therefore the search of new solvable potentials and the study of their properties is thoroughly investigated \cite{Cooper-book}--\cite{Bagrov}. At present time there are several methods to find exact solutions to the Schr\"odinger equation. We know the factorization method \cite{Schrodinger}--\cite{Infeld-Hull}, the methods based on supersymmetric quantum mechanics \cite{Cooper-book}--\cite{Bagchi-book}, \cite{Witten-susy}--\cite{Dutt-ajp-1988}, on the point canonical transformations \cite{Bhattacharjie}, and on the Darboux transformations \cite{Darboux}, \cite{Bagrov}.

Recently in Ref.\ \cite{ALO-2014-I} it is shown that for $x \in (0, +\infty)$ we can solve exactly the Schr\"odinger equations of the partner potentials
\begin{equation}\label{e: CES potentials previous}
V_\pm^I (x) = \frac{m^2}{x} \pm \frac{m}{2} \frac{1}{x^{3/2}}   ,
\end{equation}
where $m$ is a constant. We also find that these potentials are multiplicative shape invariant and each linearly independent solution includes the sum of two confluent hypergeometric functions. 

For the inverse square root potential 
\begin{equation} \label{e: square root potential}
 V_{SR} = \frac{\tilde{V}_0}{x^{1/2}},
\end{equation} 
where $\tilde{V}_0$ is a constant, in Ref.\  \cite{Ishkhanyan-1} Ishkhanyan finds that the Schr\"odinger equation is exactly solvable. Furthermore  in Ref.\ \cite{Ishkhanyan-PLA} it is shown that another exactly solvable potential is the Lambert $W$-function step potential
\begin{equation} \label{e: lambert potential}
 V_W =  \frac{\tilde{V}_0}{1 + W(\textrm{e}^{-u/\sigma})},
\end{equation} 
where $u \in (-\infty,+\infty)$, $W$ is the Lambert function, and $\sigma$ is a constant. We notice that each linearly independent solution found in Refs.\  \cite{Ishkhanyan-1}, \cite{Ishkhanyan-PLA} includes the sum of two confluent hypergeometric functions with non-constant coefficients as the exact solutions previously studied in Ref.\  \cite{ALO-2014-I}. Furthermore, in Ref.\  \cite{Ishkhanyan-2} it is shown that for the sum of the potentials (\ref{e: CES potentials previous}) and (\ref{e: square root potential}) the Schr\"odinger equation can be exactly solved and the linearly independent solutions include a sum of two confluent hypergeometric functions.

Based on the method of Ref.\  \cite{ALO-2014-I}, for the partner potentials
\begin{eqnarray} \label{e: ces step potentials}
V^{II}_\pm  = m^2 \frac{ e^u}{e^u + 1} \mp \frac{m}{2} \frac{e^{u/2}}{(e^u + 1)^{3/2}} ,
\end{eqnarray} 
where $m$ is a constant (as for the potentials (\ref{e: CES potentials previous})), in Ref.\  \cite{ALO-2015-arxiv} we showed that the Schr\"odinger equation is exactly solvable and each linearly independent solution involves a sum of two hypergeometric functions with non-constant coefficients. Thus the results of Ref.\  \cite{ALO-2015-arxiv} complement  those of Refs.\  \cite{ALO-2014-I}--\cite{Ishkhanyan-2}. Recently these results are extended in Ref.\ \cite{Ishkhanyan-last} where an extensive study is carried out of the potentials whose solutions involve Heun functions and a list of known potentials with linearly independent solutions expanded as a sum of (confluent) hypergeometric functions is given.

Our purpose in this work is to extend the results on the potentials (\ref{e: CES potentials previous}) and (\ref{e: ces step potentials}). Here we study the properties of two partner potentials defined on the half line and possessing the property that each of their linearly independent solutions includes two hypergeometric functions as those of Ref.\  \cite{ALO-2015-arxiv} and in contrast to the potentials of Refs.\ \cite{ALO-2014-I}--\cite{Ishkhanyan-2} whose linearly independent solutions include two confluent hypergeometric functions.  The expressions of the potentials that we study are 
\begin{eqnarray} \label{e: potentials ces}
V_\pm (x, m) &=&  \frac{m^2 }{e^x - 1} \pm \frac{m}{2} \frac{e^{x}}{(e^x - 1)^{3/2}} ,
\end{eqnarray} 
where, as in Refs.\  \cite{ALO-2014-I}, \cite{ALO-2015-arxiv}, $m$ is a constant. As far as we know the potentials that we present in this work are first studied. Also we notice that the potentials (\ref{e: potentials ces}) are one example of the potentials written in explicit form and whose linearly independent solutions include a sum of two hypergeometric functions with non-constant coefficients (as we show below). Previous potentials with this property appear in Ref.\  \cite{ALO-2015-arxiv} or are given in implicit form \cite{Cooper:1986tz}. 

The potentials (\ref{e: potentials ces}) are algebraic modifications of the Hulthen potential \cite{Gangopadhyayabook}
\begin{equation} \label{e: Hulthen potential}
 V_H = \frac{Q}{e^x-1},
\end{equation} 
where $Q$ is a constant. We notice that the Hulthen potential near $x=0$ behaves as $1/x$ and decays exponentially as $x \to + \infty$. It is convenient to notice that we can not obtain the Hulthen potential as a limit of the potentials (\ref{e: potentials ces}). From the shape of the potentials (\ref{e: potentials ces}) we think that they can be useful to study scattering or tunneling phenomena (see Figs.\ 2-5) \cite{Flugge}. For some values of the parameters, in an interval, one of our potentials reminds us the shape of the effective potentials that govern the propagation of the Dirac field in a Schwarzschild black hole \cite{Chandra-book}, \cite{Cho-Dirac-qnms} and therefore they can be used as a model to understand its dynamics in this spacetime. Furthermore we do not find the potentials (\ref{e: potentials ces}) in the list of Ref.\  \cite{Ishkhanyan-last} that  enumerates the known examples of potentials with linearly independent solutions involving a sum of (confluent) hypergeometric functions (see Table 3 of Ref.\  \cite{Ishkhanyan-last}). Thus we believe that these potentials are first studied in this work.

We think that the partner potentials (\ref{e: potentials ces}) may be useful in supersymmetric quantum mechanics  as a basis to generate new exactly solvable potentials of the Schrodinger equation \cite{Cooper-book}--\cite{Bagchi-book}, \cite{Fernandez}, \cite{Cooper:1994eh}. Also the mathematical form of the exact solutions is not common (see the expressions (\ref{e: Zeta 1}), (\ref{e: Zeta 2}), (\ref{e: solutions v variable})) and they can be used as a model to search new exact solutions of the Schr\"odinger equation, since exact solutions of this mathematical form appear previously in Refs.\ \cite{ALO-2014-I}--\cite{ALO-2015-arxiv}.

For several potentials we can find exact solutions to their Schr\"odinger equations in terms of special functions only when the parameters of the potentials satisfy some restrictions \cite{Bagchi-book}, \cite{Ishkhanyan-2}, \cite{SouzaDutra}--\cite{Roychoudhury}. These potentials are known as conditionally exactly solvable potentials (CES potentials in what follows). For the partner potentials (\ref{e: potentials ces}) that we study in this work we show that their parameters satisfy an algebraic constriction and therefore they are CES in the sense of Ref.\  \cite{Ishkhanyan-2}, that is, we call a potential as CES when its parameters can not be varied independently, that is, they satisfy a constriction \cite{Ishkhanyan-2}. Notice that this definition of CES potential does not impose that some parameter takes a fixed value \cite{Ishkhanyan-2}. 

We organize this paper as follows. In Sect.\ \ref{s: Method} we study the properties of the partner potentials (\ref{e: potentials ces}) that we analyze in this work. Using the method of Ref.\ \cite{ALO-2014-I} we solve exactly the Schr\"odinger equations of the studied potentials. We also expound some facts on these partner potentials and verify the solutions that we previously found.  In Sect.\ \ref{s: scattering amplitude} we calculate the scattering amplitudes of the potentials (\ref{e: potentials ces}).  We study some additional characteristics of the  potentials that we analyze in this paper in Sect.\ \ref{s: Discussion}. Finally, for the method used in this work, in Appendix we verify that it produces the solutions to the Schr\"odinger equations of partner potentials.

\section{Solution method}
\label{s: Method}

In a similar way to Refs.\ \cite{ALO-2014-I}, \cite{ALO-2015-arxiv} here we show that  in the interval $x \in (0,+\infty)$, for the partner potentials (\ref{e: potentials ces}) we can solve exactly their Schr\"odinger equations in terms of hypergeometric functions. For these partner potentials the superpotential $W$ is equal to\footnote{For the superpotential $\hat{W} = - m (\mathcal{A} e^x - \mathcal{B})^{-1/2}$ (with the constants $\mathcal{A} > \mathcal{B} > 0$), by making the change of variable $y =  x + \ln \left(\mathcal{A}/\mathcal{B} \right)$ and redefining the constant $m$ by $\hat{m}   = m / \sqrt{\mathcal{B}}$, we simplify the Schr\"odinger equations of their partner potentials to those of the potentials for the superpotential (\ref{e: superpotential ces}).}
\begin{equation} \label{e: superpotential ces}
 W (x,m) = - \frac{ m}{\sqrt{e^x - 1}} .
\end{equation} 
We notice that in Refs.\ \cite{Cooper-book}--\cite{Bagchi-book}, \cite{Ishkhanyan-last}, \cite{Khare-scattering}--\cite{Levai-search} that enumerate the solvable potentials already known, a search for the partner potentials (\ref{e: potentials ces}) shows that they have not been  previously discussed.

\begin{figure}[th]
\label{figure1}
\begin{center}
\includegraphics[scale=1,clip=true]{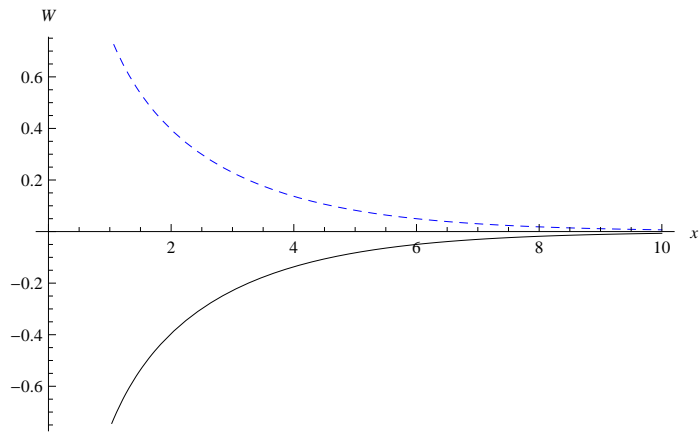}
\caption{Plots of the superpotential $W$ for $m=1$ (solid line) and for $m=-1$ (broken line).} 
\end{center}
\end{figure}

Since for the potentials (\ref{e: potentials ces}) the constants multiplying to the factors $1/(e^x -1 )$ ($m^2$) and $ e^{x} /(e^x - 1)^{3/2}$ ($ \pm m/2$) fulfill the expression $- m^2 / 4 + (\pm m /2 )^2 = 0$ these are CES potentials, as those previously studied in Refs.\  \cite{ALO-2014-I},  \cite{Ishkhanyan-2}, \cite{ALO-2015-arxiv}, \cite{SouzaDutra}--\cite{Roychoudhury}. It is convenient to notice that we classify the partner potentials (\ref{e: potentials ces}) as CES since its parameters can not be varied independently \cite{Ishkhanyan-2}. Some previously found CES potentials are \cite{Bagchi-book}, \cite{Dutt-CES}
\begin{eqnarray} \label{e: previous ces Bagchi}
 \hat{V}_1 (x) &=&  \frac{\hat{a}_1}{1 + e^{-2 u}} - \frac{\hat{b}_1}{(1 + e^{-2 u})^{1/2}} - \frac{3}{4(1 + e^{-2 u})^2} , \\
 \hat{V}_2 (x) &=& \frac{\hat{a}_2}{1 + e^{-2 u}} - \frac{\hat{b}_2 e^{-u} }{ (1 + e^{-2 u})^{1/2} }  - \frac{3}{4 (1 + e^{-2 u})^2  }, \nonumber
\end{eqnarray} 
where the constants $\hat{a}_1$ and $\hat{b}_1$ ($\hat{a}_2$ and $\hat{b}_2$) satisfy some constraints \cite{Bagchi-book}, \cite{Dutt-CES}. We point out that the CES potentials (\ref{e: previous ces Bagchi}) remind us to our potentials (\ref{e: potentials ces}), but notice that we can not get these as a limit of the CES potentials (\ref{e: previous ces Bagchi}) of Refs.\  \cite{Bagchi-book}, \cite{Dutt-CES}. Moreover the interval where they are defined is different for the CES potentials (\ref{e: potentials ces}) and (\ref{e: previous ces Bagchi}).

\begin{figure}[th]
 \begin{minipage}[b]{0.48\linewidth}
  \centering
  \includegraphics[width=\linewidth]{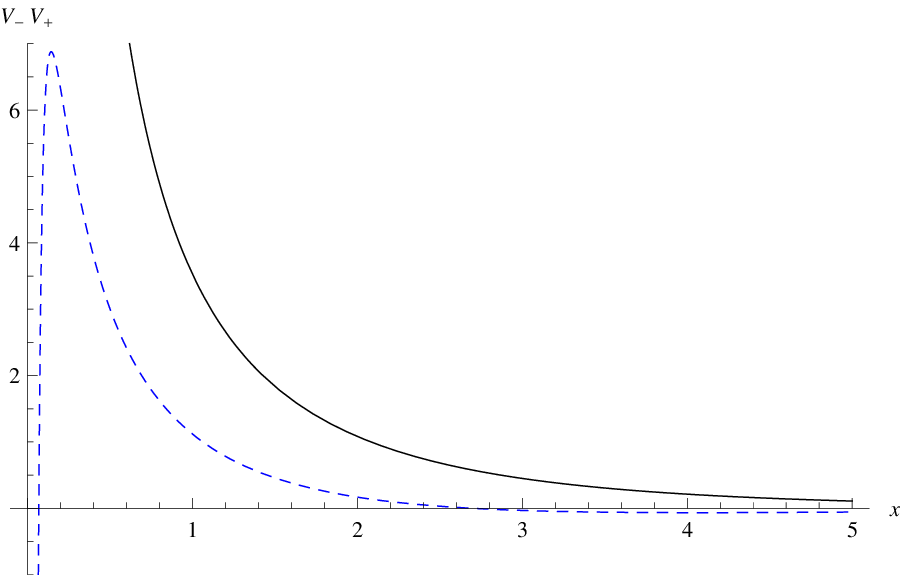}
  \caption{Plots of the potential $V_+$ (solid line) and $V_-$ (dashed line) for $m=2$.} 
  \label{figure2}
 \end{minipage}
\hspace{.02\linewidth}
\begin{minipage}[b]{0.48\linewidth}
  \centering
  \includegraphics[width=\linewidth]{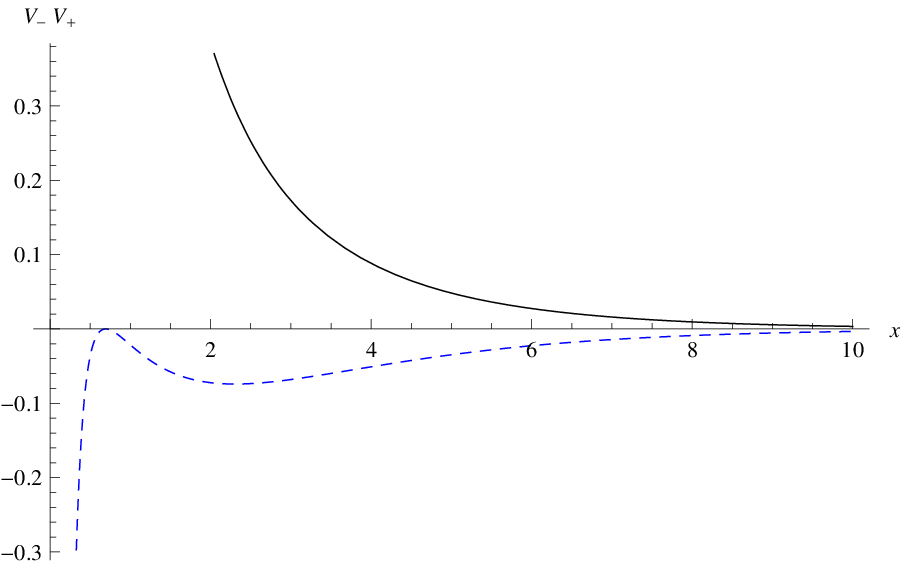}
  \caption{Plots of the potential $V_+$ (solid line) and $V_-$ (dashed line) for $m=1$.} 
  \label{figure3}
 \end{minipage}
\end{figure}

In what follows we assume that $m > 0$, since for $m < 0$ we get the same results with the potentials $V_+$ and $V_-$ interchanged. Since for $x > 0$ it is true that $1/\sqrt{e^x - 1} > 0 $, we note that the superpotential (\ref{e: superpotential ces}) does not cross the $x$-axis and therefore the supersymmetry is broken \cite{Cooper-book},  \cite{Gangopadhyayabook}. We also find
\begin{equation} \label{e: limits superpotential}
 W_+ = \lim_{x \to + \infty} W = 0^-, \qquad \qquad  \lim_{x \to 0^+} W = - \infty,
\end{equation} 
where $0^+$ ($0^-$) means that the quantity goes to zero taking positive (negative) values. Notice that as $x \to + \infty$ the superpotential $W$ decays exponentially, whereas near $x = 0$ it behaves as $1/\sqrt{x}$. Since for $m > 0$  the derivative of $W$ satisfies $\dd W /\dd x  > 0 $,
we obtain that  the superpotential is an increasing function for $x \in (0, + \infty)$. To illustrate these facts we plot the superpotential (\ref{e: superpotential ces}) in Fig.\ 1.  

For the potentials $V_{\pm}$ we get the following limits
\begin{eqnarray} \label{e: limits potentials}
 && \lim_{x \to + \infty} V_+ = 0^+, \qquad \qquad  \lim_{x \to 0^+} V_+ = + \infty ,\nonumber \\
 && \lim_{x \to + \infty} V_- = 0^-, \qquad \qquad  \lim_{x \to 0^+} V_- = - \infty .
\end{eqnarray}
Furthermore these potentials decay exponentially to zero as $x \to + \infty$ (as the Hulthen potential (\ref{e: Hulthen potential})) and near $x = 0$ they diverge as $1/ x^{3/2}$ (in a different way than the Hulthen potential (\ref{e: Hulthen potential})). We point out that near $x = 0$ the potential $V_+$ diverges to $+ \infty$, whereas the potential $V_-$ diverges to $- \infty$. The potential $V_+$ does not cross the $x$-axis and it is strictly positive, but for $m > 1$ the potential $V_-$ crosses the $x$-axis at the two points $s_{\pm} = 2 m^2 \pm 2m \sqrt{m^2-1},$ where $s = e^x $. Notice that $s_+ > s_- > 0$. For $m < 1$ the potential $V_-$ does not cross the $x$-axis and it is strictly negative.

\begin{figure}[th]
 \begin{minipage}[b]{0.48\linewidth}
  \centering
  \includegraphics[width=\linewidth]{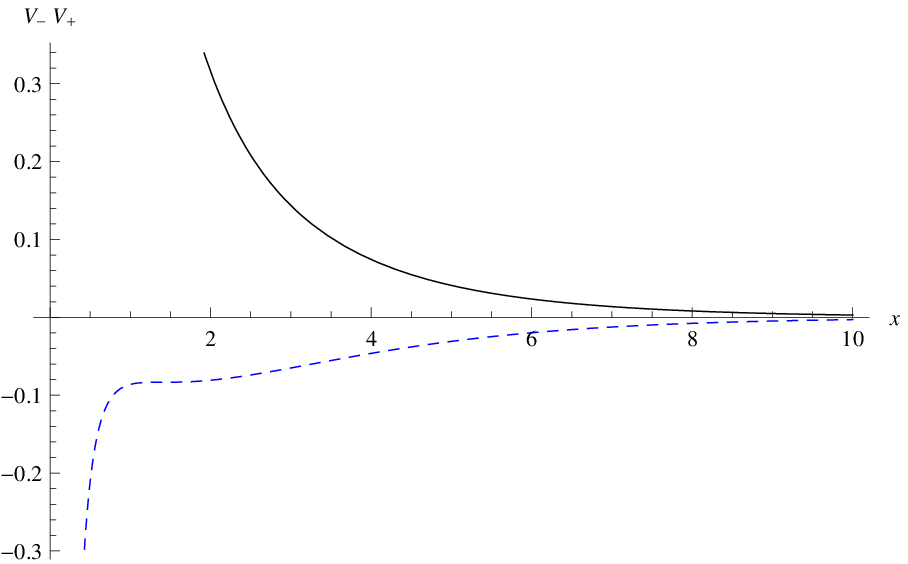}
  \caption{Plots of the potential $V_+$ (solid line) and $V_-$ (dashed line) for $m=\sqrt{3}/2$.} 
  \label{figure4}
 \end{minipage}
\hspace{.02\linewidth}
\begin{minipage}[b]{0.48\linewidth}
  \centering
  \includegraphics[width=\linewidth]{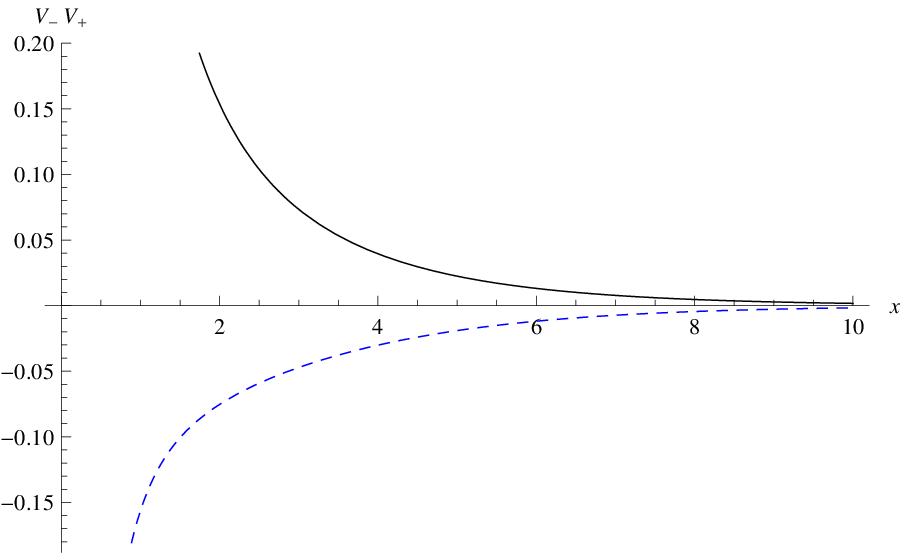}
  \caption{Plots of the potential $V_+$ (solid line) and $V_-$ (dashed line) for $m=1/2$.} 
  \label{figure5}
 \end{minipage}
\end{figure}

Owing to the derivative of the potential $V_+$ satisfies $\dd  V_+ / \dd x < 0 $, for $x \in (0,+\infty)$ we obtain that the potential $V_+$ decreases in this interval. For the potential $V_-$ we find that its derivative
\begin{equation}
 \frac{\dd  V_-}{\dd x} = - \frac{e^x}{(e^x-1)^2}\left(m^2 - \frac{m}{2} \frac{1 + e^x/2}{(e^x-1)^{1/2}} \right) ,
\end{equation} 
has critical points at $s_{1,2} = 8 m^2 - 2 \pm 4 m \sqrt{4m^2-3}$. Hence for $m > \sqrt{3}/2$ the potential $V_-$ has two real critical points, whereas for  $m < \sqrt{3}/2$ it does not have real critical points. We note that $s_1 > s_2 >0$, and we also point out that  the critical point $s_2$ is a maximum and $s_1$ is a minimum. Furthermore we notice that for $m > 1$ the quantities $s_{\pm}$ and $s_{1,2} $ satisfy $s_1 > s_+ > s_2 > s_- $, that is, the maximum of the potential $V_-$ is located between the points $s_\pm$ where $V_-$ intersects the $x$ axis and its minimum has a coordinate greater than the intersections of the potential $V_-$ with the $x$ axis. We illustrate these facts in Figs.\ 2--5.

In what follows, using the method of Ref.\ \cite{ALO-2014-I}, (see also Ref.\  \cite{ALO-2015-arxiv}) we solve exactly the Schr\"odinger equations of the partner potentials (\ref{e: potentials ces}). With this objective we write these equations as
\begin{eqnarray} \label{e: Schrodinger equations}
\frac{\dd^{2} Z_-}{\dd x^{2}} + \omega^{2}  Z_- = \left( W^2 - \frac{\dd W}{\dd x} \right) Z_- , \\
 \frac{\dd^{2} Z_+}{\dd x^{2}} + \omega^{2}  Z_+ = \left( W^2 + \frac{\dd W}{\dd x} \right) Z_+ , \nonumber 
\end{eqnarray}
and as in Ref.\ \cite{ALO-2014-I}, to simplify the equations that follow, we denote the energy $E$ as $\omega^2$. In Ref.\ \cite{ALO-2014-I} it is shown that for $\omega \neq 0$ the Schr\"odinger equations (\ref{e: Schrodinger equations}) can be written as 
\begin{eqnarray}
 \left( \frac{\dd }{\dd x } - W \right) \frac{1}{i \omega} \left( \frac{\dd }{\dd x } + W \right) Z_- = i \omega Z_-, \\
 \left( \frac{\dd }{\dd x } + W \right) \frac{1}{i \omega} \left( \frac{\dd }{\dd x } - W \right) Z_+ = i \omega Z_+, \nonumber 
\end{eqnarray}
from which we obtain that the functions $Z_+$ and $Z_-$ satisfy the coupled system
\begin{eqnarray} \label{e: Z coupled}
\left( \frac{\dd }{\dd x } + W \right) Z_- = i \omega Z_+,  \qquad  %\nonumber \\ 
  \left( \frac{\dd }{\dd x } - W \right) Z_+ = i \omega Z_- .
\end{eqnarray}

Defining $Z_{\pm} = R_1 \pm R_2, $ we get that Eqs.\  (\ref{e: Z coupled}) transform into the coupled system 
\begin{eqnarray} \label{e: equations R}
\frac{\dd R_{1} }{\dd x} - i\omega R_{1} =  W R_{2}, %\nonumber \\ 
 \qquad \qquad \frac{\dd R_{2} }{\dd x} + i\omega R_{2} = W R_{1} ,
\end{eqnarray}
(see Eqs.\ (11) of Ref.\ \cite{ALO-2014-I}). In Appendix we show that the solutions of these coupled equations produce the solutions to the Schr\"odinger equations of partner potentials.

As in Ref.\  \cite{ALO-2014-I} we take $R_1 =  e^{ - i \pi /4}  \tilde{R}_1 ,$ $R_2 = e^{i \pi /4}  \tilde{R}_2 ,$  and defining the variable $z$ by\footnote{Notice that for $x \in (0,+\infty)$ the variable $z$ varies over the range $ 0 < z < 1$.}
\begin{equation} \label{e: z definition}
 z = e^{-x},
\end{equation} 
we find that the coupled system of differential equations (\ref{e: equations R}) transforms into
\begin{eqnarray} \label{e: R tilde coupled}
 z \frac{\dd \tilde{R}_{1} }{\dd z} + i \omega \tilde{R}_{1} &=&  i m \frac{ z^{1/2}}{(1-z)^{1/2}} \tilde{R}_2 , \nonumber \\
 z \frac{\dd \tilde{R}_{2} }{\dd z} - i \omega \tilde{R}_{2} &=& -  i m \frac{ z^{1/2} }{(1-z)^{1/2}} \tilde{R}_1 .
\end{eqnarray}
From this coupled system we obtain that the functions $\tilde{R}_{1}$ and $\tilde{R}_{2}$ must be solutions of the decoupled differential equations
\begin{eqnarray} \label{e: radial equations tilde} 
\frac{\dd^{2} \tilde{R}_k }{\dd z^{2}} + \left( \frac{1/2}{z} - \frac{1/2}{1-z} \right) \frac{\dd \tilde{R}_k }{\dd z} + \frac{\omega^2- i \omega \epsilon /2  }{ z^{2} } \tilde{R}_k  - \frac{m^{2} + i \omega \epsilon /2  }{z (1-z )} \tilde{R}_k  =0 ,
\end{eqnarray} 
where $k=1,2$, and $\epsilon = 1$ ($\epsilon = -1$) for  $\tilde{R}_{1}$ ($\tilde{R}_{2}$).

If the functions $\tilde{R}_{1}$ and $\tilde{R}_{2}$ take the form $\tilde{R}_k = z^{A_k}  \bar{R}_k ,$ with the quantities $A_k$ being solutions of the algebraic equations
\begin{equation}\label{e: A equation}
 A_k^2 - \frac{A_k}{2} - \frac{i \omega \epsilon}{2} + \omega^2 = 0,
\end{equation}
we find that the functions $\bar{R}_k$ satisfy the differential equations 
\begin{equation} \label{e: equations bar R}
 \frac{\dd^{2} \bar{R}_k }{\dd z^{2}} + \left( \frac{2 A_k + 1/2}{z} - \frac{1/2}{1-z} \right) \frac{\dd \bar{R}_k }{\dd z} - \frac{m^2 +  i \omega \epsilon /2 + A_k/2 }{z(1-z)} \bar{R}_k = 0.
\end{equation} 
These equations are of hypergeometric type \cite{Abramowitz-book}--\cite{NIST-book}
\begin{equation} \label{e: hypergeometric equation}
 z(1-z)\frac{\dd^2 F}{\dd z^2} + (c -(a+b+1)z)\frac{\dd F}{\dd z} - a b F = 0 ,
\end{equation} 
with the parameters $a_k$, $b_k$, $c_k$ equal to
\begin{eqnarray} \label{e: a b c hypergeometric}
a_k = A_k + i(m^2 + \omega^2 )^{1/2}, \,\,\,\,
b_k =  A_k - i(m^2 + \omega^2 )^{1/2},  \,\,\,\, 
c_k = 2 A_k + 1/2. 
\end{eqnarray}

If the parameters $c_k$ are not integers,\footnote{We take the constants $c_k$ different from integral numbers to discard the solutions of the hypergeometric equation (\ref{e: hypergeometric equation}) that include logarithmic terms \cite{Abramowitz-book}--\cite{NIST-book}.} then the functions $\tilde{R}_{k}$ are
\begin{eqnarray} \label{e: solutions R tilde}
 \tilde{R}_k  &=&  z^{A_k}  \left[ G_k \, {}_{2}F_{1} (a_k,b_k;c_k;z) \right. \nonumber \\
  & & \left. + H_k \, z^{1-c_k}  {}_{2}F_{1}(a_k-c_k+1,b_k-c_k+1;2-c_k;z) \right] ,
\end{eqnarray}
where $\hypergf (a,b;c;z)$ denotes the hypergeometric function \cite{Abramowitz-book}--\cite{NIST-book}, and the quantities $G_k$, $H_k$ are constants.

In a straightforward way we find that Eqs.\  (\ref{e: equations R}) impose conditions on the constants $G_k$ and $H_k$. To discuss this fact we take the quantities $A_1$ and $A_2$ as $A_1 = i \omega + 1/2 = A_2 + 1/2$. Therefore the constants $a_k$, $b_k$, $c_k$ are equal to
\begin{eqnarray} \label{e: a b c hypergeometric defined}
a_1 &=& a_2 + 1/2 = 1/2 + i \omega  + i(m^2 + \omega^2 )^{1/2}, \nonumber \\
b_1 &=& b_2 + 1/2= 1/2 + i \omega  - i(m^2 + \omega^2 )^{1/2},  \\
c_1 &=& c_2 + 1 = 2 i \omega + 3/2. \nonumber
\end{eqnarray}

From Eqs.\ (\ref{e: equations R}) and the contiguous relations of the hypergeometric function \cite{Lebedev} we obtain:

a) If we choose the function $\tilde{R}_{1}$ as
\begin{equation} \label{e: R one tilde first}
 \tilde{R}_1 = G_1 z^{A_1} {}_{2}F_{1}(a_1,b_1;c_1;z),
\end{equation} 
then from Eqs.\ (\ref{e: equations R})  we get that the function $\tilde{R}_2$ must be equal to
\begin{equation} \label{e: R two tilde first}
  \tilde{R}_2 = G_1 \frac{c_1 - 1}{im}   z^{A_2} {}_{2}F_{1}(a_2,b_2;c_2;z),
\end{equation} 
and the constants $G_1$ and $G_2$ are related by $G_2 = G_1 (c_1 - 1)/ (im ) $.

b) If we select the function $\tilde{R}_{1}$ in the form
\begin{equation}
 \tilde{R}_1  = H_1  z^{A_1 + 1-c_1}  {}_{2}F_{1}(a_1-c_1+1,b_1-c_1+1;2-c_1;z) ,
\end{equation} 
then from Eqs.\  (\ref{e: equations R}) we obtain that  the function $\tilde{R}_2$ must be equal to
\begin{eqnarray}
 \tilde{R}_2 &=& H_1  \frac{ (a_1-c_1+1)(b_1-c_1+1)}{im(2-c_1)}  z^{A_2 + 1-c_2} \nonumber \\  
 & & \times {}_{2}F_{1}(a_2-c_2+1,b_2-c_2+1;2-c_2;z) ,
\end{eqnarray} 
and the constants $H_1$ and $H_2$ satisfy $H_2 = H_1  (a_1-c_1+1)(b_1-c_1+1)/(im(2-c_1))$. Hence we find that Eqs.\  (\ref{e: equations R}) impose the previous constrictions on the constants $G_k$ and $H_k$.

Considering the previous definitions we get that as function of $\tilde{R}_1$ and $\tilde{R}_2$ the solutions $Z_{\pm}$ take the form $Z_\pm = e^{-i \pi /4} (\tilde{R}_1 \pm i \tilde{R}_2) $. Thus from our results we get that the linearly independent solutions to the Schr\"odinger equations of the potentials $V_\pm$ are
\begin{equation}\label{e: Zeta 1}
 Z_\pm^I = G_1 \textrm{e}^{-i \pi /4}  \left( z^{A_1} \hypergf(a_1,b_1;c_1;z) \pm \frac{c_1-1}{m} z^{A_2} \hypergf(a_2,b_2;c_2;z) \right),
\end{equation}
and
\begin{eqnarray}\label{e: Zeta 2}
 Z_\pm^{II} &=& H_1 \textrm{e}^{-i \pi /4}  \left( z^{A_1+1-c_1} \hypergf(a_1-c_1+1,b_1-c_1+1;2-c_1;z) \right.  \\
  &&  \pm z^{A_2+1-c_2}   \frac{(a_1-c_1+1)(b_1-c_1+1)}{(2-c_1) m}   \nonumber  \\
 & &\times 
 \left. \hypergf(a_2-c_2+1,b_2-c_2+1;2-c_2;z) \right). \nonumber
\end{eqnarray}

Using that for the linearly independent solutions to the hypergeometric differential equation (\ref{e: hypergeometric equation}) its Wronskian is \cite{NIST-book}
\begin{equation}
 \tilde{W}_z [ \hypergf (a,b;c;z), z^{1-c} \hypergf (a-c+1, b-c+1; 2-c;z) ]  = \frac{(1-c)}{ z^{c} (1-z)^{a+b+1-c}}, 
\end{equation} 
in a straightforward way we find that the Wronskian of the solutions $Z_\pm^I$ and $Z_\pm^{II}$ is equal to (for $G_1 = H_1 = 1$)
\begin{equation}
 \mathfrak{W}_x [Z_\pm^I,Z_\pm^{II}] = \pm \frac{2 \omega (c_1-1)}{m} .
\end{equation} 

Furthermore, taking into account Eqs.\ (\ref{e: equations bar R}), we get that the functions $\tilde{R}_{k}$ satisfy
\begin{eqnarray} \label{e: second derivative tilde R}
 \frac{\dd }{\dd z} \left( z \frac{\dd \tilde{R}_k}{\dd z} \right) &=& \frac{1/2}{1-z} \frac{\dd \tilde{R}_k }{\dd z} - \frac{A_k}{2(1-z)} \tilde{R}_k \nonumber \\
  & +& \frac{A_k^2 - A_k/2}{z} \tilde{R}_k +   \frac{m^2 + A_k/2 + i \epsilon \omega/2}{1-z}   \tilde{R}_k.
\end{eqnarray} 
From these equations we get that the functions $Z_\pm$ fulfill 
\begin{equation} \label{e: Schrodinger equation z variable}
 \frac{\dd }{\dd z} \left( z \frac{\dd Z_\pm}{\dd z}  \right) + \left( \frac{\omega^2}{z} - \frac{m^2}{1-z} \mp \frac{m}{2} \frac{1}{z^{1/2}(1-z)^{3/2}} \right) Z_\pm = 0,
\end{equation} 
that are the Schr\"odinger equations (\ref{e: Schrodinger equations}) in the  variable $z$ defined in the expression (\ref{e: z definition}). Hence, if the functions $\bar{R}_k$ are solutions of Eqs.\  (\ref{e: equations bar R}), then the functions $Z_{\pm}$ solve the Schr\"odinger equations (\ref{e: Schrodinger equations}) with the potentials (\ref{e: potentials ces}).

%\newpage

\section{Scattering amplitude}
\label{s: scattering amplitude}

In what follows we determine the scattering amplitudes for the potentials $V_\pm$. To calculate the scattering amplitude it is convenient to use the variable $v=1-z$ to write the solutions of the Schr\"odinger equations.\footnote{We point out that the variable $v$ varies over the range $0 < v < 1$.} We find that in this variable,  the linearly independent solutions of the Schr\"odinger equations for the potentials (\ref{e: potentials ces}) take the form
\begin{eqnarray} \label{e: solutions v variable}
 \tilde{Z}_\pm^I &=& \tilde{C}_1 \left( (1-v)^{B_1} \hypergf (\alpha_1,\beta_1;\gamma_1;v) \right. \nonumber \\ 
 &-& \left.  \frac{m}{\gamma_1} (1-v)^{B_2} v^{1-\gamma_2} \hypergf (\alpha_2-\gamma_2+1,\beta_2-\gamma_2+1;2-\gamma_2;v) \right) ,\nonumber  \\
 \tilde{Z}_\pm^{II} &=& \tilde{C}_2 \left(  (1-v)^{B_1} v^{1-\gamma_1} \hypergf (\alpha_1-\gamma_1+1,\beta_1-\gamma_1+1;2-\gamma_1;v) \nonumber \right. \\
 &-& \left. \frac{\gamma_1}{m} (1-v)^{B_2} \hypergf (\alpha_2,\beta_2;\gamma_2;v) \right) ,
\end{eqnarray} 
where $\tilde{C}_1$, $\tilde{C}_2$ are constants and 
\begin{eqnarray} \label{e: alpha beta gamma hypergeometric}
B_1 &=& 1/2 + i \omega, \,\,\,\,\,\,\,\,\,\, \quad B_2 = i \omega , \nonumber \\
\alpha_k &=& B_k + i(m^2 + \omega^2 )^{1/2}, \,\,\,\,
\beta_k =  B_k - i(m^2 + \omega^2 )^{1/2},  \,\,\,\, 
\gamma_k = 1/2. 
\end{eqnarray}

In what follows we study in detail the potential $V_+$ since a similar calculation produces the result for the potential $V_-$. We see that near $x=0$ ($v=0$) the solutions $\tilde{Z}_\pm$ behave as
\begin{equation}
 \tilde{Z}_\pm^I \approx 1 - \frac{m}{\gamma_1} v^{1/2} , \qquad \qquad  \tilde{Z}_\pm^{II} \approx  - \frac{\gamma_1}{m} + v^{1/2} .
\end{equation} 
Since we like to impose as boundary condition that the solution is equal to zero at $x=0$ it is convenient to define the new solutions 
\begin{equation}
 Y_+^I = \tilde{Z}_\pm^I - \frac{m}{\gamma_1} \tilde{Z}_\pm^{II} ,  \qquad \qquad   Y_+^{II} =  \tilde{Z}_\pm^I + \frac{m}{\gamma_1} \tilde{Z}_\pm^{II},
\end{equation} 
that near $x=0$  behave as
\begin{equation}
 Y_+^I \approx 2 -  \frac{2 m}{\gamma_1} v^{1/2} , \qquad \qquad Y_+^{II} \approx  0.  
\end{equation} 
Therefore to satisfy the boundary condition we choose the solution $Y_+^{II}$ that is equal to zero at $x=0$. 

Taking into account the Kummer property of the hypergeometric function \cite{Abramowitz-book}--\cite{NIST-book}
\begin{eqnarray} \label{e: Kummer property v 1-v}
& &{}_2F_1(a,b;c;v) = \frac{\Gamma(c) \Gamma(c-a-b)}{\Gamma(c-a) \Gamma(c - b)} {}_2 F_1 (a,b;a +b +1-c;1-v)  \\
&+& \frac{\Gamma(c) \Gamma( a + b - c)}{\Gamma(a) \Gamma(b)} (1-v)^{c-a-b} {}_2F_1(c-a, c-b; c + 1 -a-b; 1 -v) , \nonumber
\end{eqnarray}
we obtain that as $x \to \infty$ ($v \to 1$) the solution $Y_+^{II}$ behaves as
\begin{eqnarray}
 Y_+^{II} &\approx& \frac{\Gamma(1/2 + 2 i \omega) 2^{8 i \omega}}{\Gamma(1/2 - 2 i \omega)} \frac{ \Gamma(-2 \alpha_2)\Gamma(-2 \beta_2) }{\Gamma(2 \alpha_2)\Gamma(2 \beta_2)} \\
 &\times& \frac{m \Gamma( \alpha_2) \Gamma( \beta_2) +\Gamma( 1/2+\alpha_2) \Gamma(1/2+ \beta_2)}{m \Gamma(- \alpha_2) \Gamma(- \beta_2) +\Gamma( 1/2 -\alpha_2) \Gamma(1/2 - \beta_2)} \textrm{e}^{i \omega x} - \textrm{e}^{- i \omega x} , \nonumber 
\end{eqnarray}
that is, the scattering amplitude of the potential $V_+$ is equal to \cite{Cooper-book}, \cite{Flugge}
\begin{eqnarray}
 S_+ &=& \frac{\Gamma(1/2 + 2 i \omega) 2^{8 i \omega}}{\Gamma(1/2 - 2 i \omega)} \frac{ \Gamma(-2 \alpha_2)\Gamma(-2 \beta_2) }{\Gamma(2 \alpha_2)\Gamma(2 \beta_2)} \nonumber \\
 &\times& \frac{m \Gamma( \alpha_2) \Gamma( \beta_2) +\Gamma( 1/2+\alpha_2) \Gamma(1/2+ \beta_2)}{m \Gamma(- \alpha_2) \Gamma(- \beta_2) +\Gamma( 1/2 -\alpha_2) \Gamma(1/2 - \beta_2)} .
\end{eqnarray} 
Notice that the previous scattering amplitude satisfies $S_+ S_+^* = 1 $. A similar result is valid for the scattering amplitude of the potential $V_-$.

%\newpage

\section{Discussion}
\label{s: Discussion}

Here we show that each of the exact solutions (\ref{e: Zeta 1}) and (\ref{e: Zeta 2})  of the Schr\"odinger equations for the partner potentials (\ref{e: potentials ces}) includes a sum with non-constant coefficients of two hypergeometric functions (see also the exact solutions (\ref{e: solutions v variable})). We note that this form of the solutions is not common in the previous references \cite{Cooper-book}--\cite{Bagchi-book}, \cite{Cooper:1994eh}. The potentials that we study in this work may be suitable to analyze  scattering and tunneling phenomena and they can be taken as a basis to search new exactly solvable potentials, since the mathematical form of their solutions is not widely explored. 

To finish this work we notice the following facts on the potentials (\ref{e: potentials ces}).

\begin{itemize}

\item Considering that $ e^{x} = e^{x/2} / e^{-x/2}$ and employing hyperbolic functions we obtain that the superpotential (\ref{e: superpotential ces}) and the potentials (\ref{e: potentials ces}) take the form
\begin{eqnarray}
 W &=& - \mu (\coth(x/2)-1)^{1/2}, \nonumber \\
 V_\pm &=& \mu^2  (\coth(x/2) -1) \pm \frac{\mu}{4} \frac{(1 + \coth(x/2) )^{1/2}}{\sinh(x/2)} ,
\end{eqnarray}
with $\mu = m / \sqrt{2}$.

\item We notice that near $x = 0$, the potentials $V_\pm$ behave as (preserving the leading and subleading terms)
\begin{equation} \label{e: behavior near zero}
 \frac{m^2}{x} \pm \frac{m}{2} \frac{1}{x^{3/2}} ,
\end{equation} 
that are the potentials (\ref{e: CES potentials previous}) previously studied in Ref.\  \cite{ALO-2014-I}, that is, near $x=0$ our potentials $V_\pm$ yield the behavior analyzed in Ref.\ \cite{ALO-2014-I}. In contrast to the potentials of Ref.\  \cite{ALO-2014-I}, the potentials $V_\pm$ decay exponentially as $x \to + \infty$ (the potentials (\ref{e: CES potentials previous}) decay as $1/x$ as $x \to + \infty$). Thus we can consider to the potentials (\ref{e: potentials ces}) as a generalization of the potentials (\ref{e: CES potentials previous}).

\item The partner potentials are shape invariant if they satisfy $V_+(x,\alpha_0) = V_-(x,\alpha_1) + R (\alpha_0) $ \cite{Gendenshtein},  where the parameters $\alpha_0$, $\alpha_1$ are independent of the coordinate $x$, with $\alpha_1 = f(\alpha_0)$, and $R (\alpha_0)$ is also a function of $\alpha_0$. From the expressions (\ref{e: potentials ces}) we notice that the potentials  $V_\pm$ fulfill $V_-(x,-m)   =  V_+(x,m),$ and therefore they are multiplicative shape invariant, since $\alpha_0 = m$, $\alpha_1 = - \alpha_0 = q \alpha_0$ with $q=-1$ and $R (\alpha_0) = 0$. For several multiplicative shape invariant potentials that are already found \cite{Cooper-book}, \cite{Cooper:1994eh}, we know them in series form, but we get the potentials (\ref{e: potentials ces}) in closed form, in a similar way to the multiplicative shape invariant potentials of Refs.\ \cite{ALO-2014-I}, \cite{ALO-2015-arxiv}.

Recently in Ref.\ \cite{Aleixo-Balantekin} is studied the  concept of shape invariance with reflection transformations. The analyzed transformations include reflections of the coordinates and translations of the parameters. For the potentials (\ref{e: potentials ces}) the formula $\alpha_1 = - \alpha_0 $ remind us a reflection, but for the parameters of the potential and it is different from the mathematical operations considered in Ref.\  \cite{Aleixo-Balantekin}. 

\item As previously noted, the superpotential $W$ does not cross the $x$-axis and therefore the supersymmetry is broken \cite{Cooper-book}, \cite{Gangopadhyayabook}. Thus the functions $\psi^{\mp}_0$ that are solutions of the differential equations
\begin{equation}
 \left(\frac{\dd }{\dd x} + W \right) \psi_0^- = 0, \qquad \left(- \frac{\dd }{\dd x} + W \right) \psi_0^+ = 0,
\end{equation} 
are equal to
\begin{equation}
 \psi^{\mp}_0 = \exp \left( \mp \int W(x^{\prime}) \dd x^{\prime} \right) = \left[ \sqrt{1-\textrm{e}^{-x}} + i \textrm{e}^{-x/2} \right]^{\pm 2 i m} ,
\end{equation} 
and they are not normalizable. 

Taking into account the formulas (15.4.11) and (15.4.15) of Ref.\  \cite{NIST-book}, we obtain that the functions $\psi^{\mp}_0$ can be written in the form
\begin{equation} \label{e: psi 0 1 hypergeometric}
 \psi^{\mp}_0 = \hypergf (\mp im, \pm im; 1/2;z) \mp 2 m z^{1/2} \hypergf (1/2 \mp im, 1/2 \pm i m; 3/2; z) .
\end{equation} 
From the values of the parameters $A_k$ and $a_k$, $b_k$, and $c_k$ with $\omega = 0$ we find that the functions $\tilde{R}_k$ of the expressions (\ref{e: R one tilde first}) and (\ref{e: R two tilde first}) simplify to
\begin{eqnarray} \label{e: tilde for omega 0}
 \tilde{R}_1 &=& z^{1/2} \hypergf   (1/2 + im, 1/2 - i m; 3/2; z) , \nonumber \\
 \tilde{R}_2 &=& \frac{1}{2mi} \hypergf (im, - im; 1/2;z).
\end{eqnarray}
Thus from the expressions (\ref{e: psi 0 1 hypergeometric}) and (\ref{e: tilde for omega 0}) we obtain
\begin{equation}
 \psi^{\mp}_0 = \mp 2 m (\tilde{R}_1 \mp i \tilde{R}_2) = \mp 2 m \textrm{e}^{i \pi /4} Z_\mp.
\end{equation} 
Hence the functions $\psi^{\mp}_0$ are proportional to $Z_\mp$, as we expect from the previous analysis. 

\item Finally we notice that for the potentials (\ref{e: potentials ces}) each linearly independent solution of the Schr\"odinger equations (\ref{e: Schrodinger equations}) includes a sum with non-constant coefficients of two hypergeometric functions (see the expressions (\ref{e: Zeta 1}),  (\ref{e: Zeta 2}), and (\ref{e: solutions v variable})), and we have not been able to simplify this sum to a single hypergeometric function, but this fact must be studied carefully. Therefore, as those of Refs.\ \cite{ALO-2014-I}--\cite{ALO-2015-arxiv}, the potentials (\ref{e: potentials ces}) are examples of the potentials analyzed, but not given in explicit form in Ref.\ \cite{Cooper:1986tz} whose linearly independent solutions include a sum of (confluent) hypergeometric functions. 

From the results of Ref.\  \cite{Ishkhanyan-last} we think that the solutions of the Schr\"odinger equations for the potentials (\ref{e: potentials ces}) also can be expanded in terms of Heun functions. As far as we can see the advantage of writing the solutions (\ref{e: Zeta 1}) and (\ref{e: Zeta 2}) (see also (\ref{e: solutions v variable})) as a sum of hypergeometric functions (instead of Heun functions) is that we can use the well developed techniques involving (confluent) hypergeometric functions (as Kummer's property (\ref{e: Kummer property v 1-v})) in the study of the characteristics for the potentials (\ref{e: potentials ces}), as illustrated in Sect.\ \ref{s: scattering amplitude} (and in Refs.\  \cite{Ishkhanyan-PLA}, \cite{ALO-2015-arxiv}). Thus we think that it is convenient to search and study the potentials whose linearly independent solutions have this property.

To generalize the results of Refs.\ \cite{ALO-2014-I}--\cite{ALO-2015-arxiv} and this paper, a problem to analyze in detail is the search of potentials with the property that each linearly independent solution includes a sum of three or more (confluent) hypergeometric functions.

%\item 
\end{itemize}

\section{Appendix}
\label{Appendix}

In this Appendix we show that the solutions to the system of coupled equations (\ref{e: equations R}) produce the solutions to the Schr\"odinger equations of the partner potentials
\begin{equation}\label{e: partner potentials}
 \hat{V}_\pm =  W^2 \pm \frac{\dd W}{\dd x}.
\end{equation} 
First, from Eqs.\ (\ref{e: Schrodinger equations}) and the definition of $Z_\mp$  we notice that the Schr\"odinger equations for the partner potentials can be written in the form
\begin{equation}
 \frac{\dd}{\dd x} \frac{\dd}{\dd x}  (R_1 \pm R_2) + \omega^2 (R_1 \pm R_2) - \left(W^2 \pm \frac{\dd W}{\dd x} \right) (R_1 \pm R_2) = 0.
\end{equation} 
Using Eqs.\ (\ref{e: equations R}) we get that the left hand sides of the previous equations transform into 
\begin{eqnarray} \label{e: equations appendix second}
  \frac{\dd}{\dd x} \left( i \omega R_1 \mp i \omega R_2 + W R_2 \pm W R_1 \right) &+& \omega^2 (R_1 \pm R_2) \\
  &-& \left(W^2 \pm \frac{\dd W}{\dd x} \right) (R_1 \pm R_2) . \nonumber
\end{eqnarray} 

Expanding the first four factors of the previous expressions and considering Eqs.\ (\ref{e: equations R}) we find that the formulas (\ref{e: equations appendix second}) become 
\begin{eqnarray}
 &-& \omega^2 R_1 \mp \omega^2 R_2 + \frac{\dd W}{\dd x} (R_2 \pm R_1) + W^2 R_1 \pm W^2 R_2 \nonumber \\
 &+& \omega^2 (R_1 \pm R_2) - \left(W^2 \pm \frac{\dd W}{\dd x} \right)  (R_1 \pm R_2) .
\end{eqnarray}
Simplifying the previous expressions we obtain
\begin{equation}
  \frac{\dd W}{\dd x} (R_2 \pm R_1)  - \frac{\dd W}{\dd x} (R_2 \pm R_1)  = 0. 
\end{equation} 
Therefore from the solutions of the coupled system (\ref{e: equations R}) we obtain the solutions to the Schr\"odinger equations of the partner potentials (\ref{e: partner potentials}).

\section{Acknowledgments}

I thank the support by CONACYT M\'exico, SNI M\'exico, EDI-IPN, COFAA-IPN, and Research Project IPN SIP-20160074.


\begin{thebibliography}{}


\bibitem{Cooper-book}
Cooper F, Khare A, Sukhatme U 2001 {\it Supersymmetry in Quantum Mechanics} (Singapore: World Scientific)

\bibitem{Gangopadhyayabook}
Gangopadhyaya A, Mallow J V, Rasinariu C 2011 {\it Supersymmetric Quantum Mechanics: An Introduction} (Singapore: World Scientific)

\bibitem{Bagchi-book}
Bagchi B K 2000 {\it Supersymmetry in Quantum and Classical Physics} (Boca Raton: Chapman and Hall/CRC Press)   

\bibitem{Schrodinger}
Schr\"odinger E 1940 Proc.\ R.\ Ir.\ Acad.\ A \textbf{46} 183 

\bibitem{Schrodinger2}
Schr\"odinger E 1941 Proc.\ R.\ Ir.\ Acad.\ A \textbf{47} 53  

\bibitem{Infeld-Hull}
Infeld L, Hull T E 1951 Rev.\ Mod.\ Phys.\ \textbf{23} 21

\bibitem{Witten-susy}
Witten E 1981 Nucl.\ Phys.\ B \textbf{188} 513  

\bibitem{sukumar}
Sukumar C V 1985 J.\ Phys.\ A: Math.\ Gen.\ \textbf{18} 2917  

\bibitem{Fernandez}
Fern\'andez D J and Fern\'andez-Garc\'{\i}a N 2005 AIP Conf.\ Procc.\  \textbf{744} 236. {\it Supersymmetries in Physics and Its Applications}, edited by Bijker R, et al.\ (American Institute of Physics)

\bibitem{Cooper:1994eh}
Cooper F, Khare A, Sukhatme U 1995 Phys.\ Rep.\ \textbf{251} 267  

\bibitem{Dutt-ajp-1988}
Dutt R, Khare A, Sukhatme U P 1988 Am.\ J.\ Phys.\ \textbf{56} 163

\bibitem{Bhattacharjie}
Bhattacharjie A and Sudarshan E C G 1962 Nuovo Cimento \textbf{25} 864

\bibitem{Darboux}
Darboux G 1888 {\it Th\'eorie G\'en\'erale des Surfaces.\ Vol.\ 2} (Paris: Gauthier-Villars)  

\bibitem{Bagrov}
Bagrov V G and Samsonov B F 1995 Theor.\ Math.\ Phys.\ \textbf{104} 1051  

\bibitem{ALO-2014-I} 
L\'opez-Ortega A 2015 Phys.\ Scr.\ \textbf{90} 085202

\bibitem{Ishkhanyan-1}
Ishkhanyan A M 2015 Eur.\ Phys.\ Lett.\ \textbf{112}  10006

\bibitem{Ishkhanyan-PLA}
Ishkhanyan A M 2016 Phys. Lett. A \textbf{380} 640

\bibitem{Ishkhanyan-2}
Ishkhanyan A M, arXiv:1511.03565. A conditionally exactly solvable generalization of the inverse square root potential

\bibitem{ALO-2015-arxiv}  
L\'opez-Ortega A, arXiv:1512.04196. A conditionally exactly solvable generalization of the potential step

\bibitem{Ishkhanyan-last}
Ishkhanyan A M, arXiv:1601.03360. Schr\"odinger potentials solvable in terms of general Heun functions 

\bibitem{Cooper:1986tz} 
Cooper F, Ginocchio J N, Khare A 1987 Phys.\ Rev.\ D {\bf 36} 2458

\bibitem{Flugge} 
Fl\"ugge S 1971 {\it Practical Quantum Mechanics I, II} (Berlin: Springer Verlag)

\bibitem{Chandra-book}
Chandrasekhar S 1983 {\it The Mathematical Theory of Black Holes} (Oxford, Oxford University Press)

\bibitem{Cho-Dirac-qnms} 
Cho H T 2003 Phys.\ Rev.\ D  \textbf{68} 024003 

\bibitem{SouzaDutra}
de Souza Dutra A 1993 Phys.\ Rev.\ A \textbf{47} R2435

\bibitem{Dutt-CES}
Dutt R, Khare A, Varshni Y P 1995 J.\ Phys.\ A: Math.\ Gen.\ \textbf{28} L107

\bibitem{Roychoudhury}
Roychoudhury R, Roy P, Znojil M,  Levai G 2001  J.\ Math.\ Phys.\ \textbf{42} 1996

\bibitem{Khare-scattering}
Khare A, Sukhatme U P 1988 J.\ Phys.\ A: Math.\ Gen.\ \textbf{21} L501

\bibitem{Derezinski:2010ku} 
Derezinski J, Wrochna M 2011 Annales Henri Poincare {\bf 12} 397

\bibitem{Dabrowska}
Dabrowska J W,  Khare A, Sukhatme U P 1988 J.\ Phys.\ A: Math.\ Gen.\ \textbf{21} L195 

\bibitem{levai-su-algebra}
Levai G 1994 J.\ Phys.\ A: Math.\ Gen.\ \textbf{27} 3809

\bibitem{Levai-search}
Levai G 1989 J.\ Phys.\ A: Math.\ Gen.\ \textbf{22} 689

\bibitem{Abramowitz-book} 
Abramowitz M, Stegun I A 1965 {\it Handbook of Mathematical Functions with Formulas, Graphs, and Mathematical Tables}  (New York: Dover Publications)

\bibitem{Lebedev} 
Lebedev N N 1972 {\it Special Functions and Their Applications,} (New York, Dover Publications)

\bibitem{Guo-book} 
Wang Z X, Guo D R 1989 {\it Special Functions} (Singapore: World Scientific)

\bibitem{NIST-book} 
Olver F W J, Lozier D W, Boisvert R F, Clark C W 2010 {\it NIST Handbook of Mathematical Functions} (Cambridge: Cambridge University Press)

\bibitem{Gendenshtein} 
Gendenshtein L E 1983 JETP Lett.\ \textbf{38} 356 

\bibitem{Aleixo-Balantekin}  
Aleixo A N F, Balantekin A B 2014 J.\ Phys.\ A: Math.\ Theor.\ \textbf{47} 135304


\end{thebibliography}
\end{document}